\documentclass[aps,prl,reprint,amsmath]{revtex4-1}
\usepackage{bm}
\usepackage{graphicx}
\usepackage{braket} 
\usepackage{hyperref}

\newcommand{\vect}[1]{\boldsymbol{#1}}
\newcommand{\hk}{\hat{k}}
\newcommand{\hvk}{\hat{\vect{k}}}

\newcommand{\hx}{\hat{\vect{x}}}
\newcommand{\hy}{\hat{\vect{y}}}
\newcommand{\hz}{\hat{\vect{z}}}
\newcommand{\ha}{\hat{\vect{a}}}
\newcommand{\hc}{\hat{\vect{c}}}
\newcommand{\vk}{\vect{k}}
\newcommand{\vd}{\vect{d}}
\newcommand{\hB}{\hat{\vect{B}}}

\begin{document}

\title{Is the anisotropy of the upper critical field of Sr$_2$RuO$_4$ consistent with  a helical $p$-wave state?}

\author{Jingchuan Zhang}
\affiliation{Department of Physics, University of Science and Technology Beijing, Beijing 100083, China}
\author{Christopher L\"{o}rscher }
\affiliation{Department of Physics, University of Central Florida, Orlando, FL 32816-2385 USA}
\author{Qiang Gu}
\affiliation{Department of Physics, University of Science and Technology Beijing, Beijing 100083, China}
\author{Richard A. Klemm}
\affiliation{Department of Physics, University of Central Florida, Orlando, FL 32816-2385 USA}


\date{\today}

\begin{abstract}
	We calculate the angular and temperature $T$ dependencies of the upper critical field $H_{c2}(\theta,\phi,T)$ for the $C_{4v}$ point group helical $p$-wave states,  assuming a single uniaxial ellipsoidal Fermi surface, Pauli limiting, and strong spin-orbit coupling that locks the spin-triplet ${\bm d}$-vectors onto the layers.   Good fits to  the Sr$_2$RuO$_4$ $H_{c2,a}(\theta,T)$ data of Kittaka {\it et al.} [2009 {\it Phys. Rev.} B {\bf 80}, 174514] are obtained. Helical states with ${\bm d}({\bm k})=\hk_x\hx-\hk_y\hy$ and $\hk_y\hx+\hk_x\hy$ (or $\hk_x\hx+\hk_y\hy$ and $\hk_y\hx-\hk_x\hy$) produce  $H_{c2}(90^{\circ},\phi,T)$ that greatly exceed (or do not exhibit) the  four-fold azimuthal anisotropy magnitudes observed in Sr$_2$RuO$_4$ by Kittaka {\it et al.} and by Mao {\it et al.} [2000 {\it Phys. Rev. Lett.} {\bf 84} 991], respectively.
\end{abstract}
\maketitle

\section{Introduction}
Despite two decades of extensive studies the detailed structure of the superconducting order parameter in Sr$_2$RuO$_4$ remains unclear~\cite{Mackenzie:2003zz, Sigrist:2005gr, Maeno:2012ew}. Nuclear magnetic resonance (NMR) and nuclear quadruple resonance (NQR) Knight shift measurements of the electronic spin susceptibility of the O~\cite{Ishida:1998es,Mukuda:1999ep} and Ru~\cite{Ishida:2001ei, Murakawa:2004ha, Murakawa:2007ht} nuclear sites, internal magnetic field measurements by spin-polarized neutron scattering~\cite{Riseman:1998cz, Duffy:2000gk} and spin-relaxation measurements by muon spin resonance ($\mu$SR)~\cite{Luke:1998bo} all provided support to a parallel-spin pairing state. The invariance of the spin susceptibility on entering the superconducting state with the magnetic field ${\bm H}$  both parallel and perpendicular to the RuO$_2$ layers was argued to be consistent with very weak spin-orbit coupling in Sr$_2$RuO$_4$, so that the $d$-vector representing the orientation of the spin-triplet pairing state would always rotate to be perpendicular to ${\bm H}$  for $\mu_0H > 20$~mT~\cite{Murakawa:2004ha}, where $\mu_0$ is the vacuum magnetic permeability.  However, this scenario is in direct conflict with the suppression of the in-plane upper critical field $H_{c2,ab}$ ($\sim 1.5$~T) at low temperatures $T$~\cite{Deguchi:2002bv, Kittaka:2009hs, Yonezawa:2013je}, reminiscent of the strong Pauli pairbreaking limit  in spin-singlet pair states~\cite{Clogston:1962cm} or a spin-triplet pair state with the $d$-vector parallel to the field~\cite{Leggett:1975jf}. Indeed, with the assumption that the $d$-vector is locked in some direction in the basal plane, the suppression of $H_{c2,ab}$ could possibly be explained by the inclusion of Pauli pairbreaking~\cite{Machida:2008cd, Choi:2010ea}. The discrepancies in the orientation of the $d$-vector are even aggravated by the extreme sensitivity of the $H_{c2,ab}$ suppression~\cite{Kittaka:2009hs} as well as by the in-plane anisotropy of $H_{c2,ab}(\phi)$~\cite{Mao:2000gn,Deguchi:2004dk} to the precise field alignment. Although introducing a multi-component order parameter seems rather unconvincing that it might apply to all cases~\cite{Agterberg:2001hh, Mineev:2008ki}, it might be relevant to the chiral-nonchiral transition in vortex states~\cite{Ishihara:2013jb, Zhang:2014ul} or even to the first-order transition to the normal state~\cite{Yonezawa:2013je}. Further complicating matters, one set of scanning tunneling microscopy (STM) experiments was consistent with a single nodeless gap on all three Fermi surfaces of Sr$_2$RuO$_4$~\cite{Suderow}, but in another STM experiment, the tip was placed in a spot with   substantial normal regions for $T\ll T_c$~\cite{Firmo}, completely disguising any possible superconducting order parameter form.  To gain a possibly consistent interpretation to all pieces of experimental evidence, it appears indispensable to introduce a new mechanism to describe the nontrivial interaction between spin-triplet superconductivity and ${\bm H}$. Beforehand, one could nevertheless assume that the Pauli limit was essential to determine the in-plane $H_{c2,ab}$.  In addition, since many examples of anomalous Knight shift results in singlet-spin layered and heavy fermion superconductors have been obtained, a new theory of the Knight shift is sorely needed~\cite{Mazin,HallKlemm}.

\begin{figure}[htbp]
	\centering
	\includegraphics[width=0.45\textwidth]{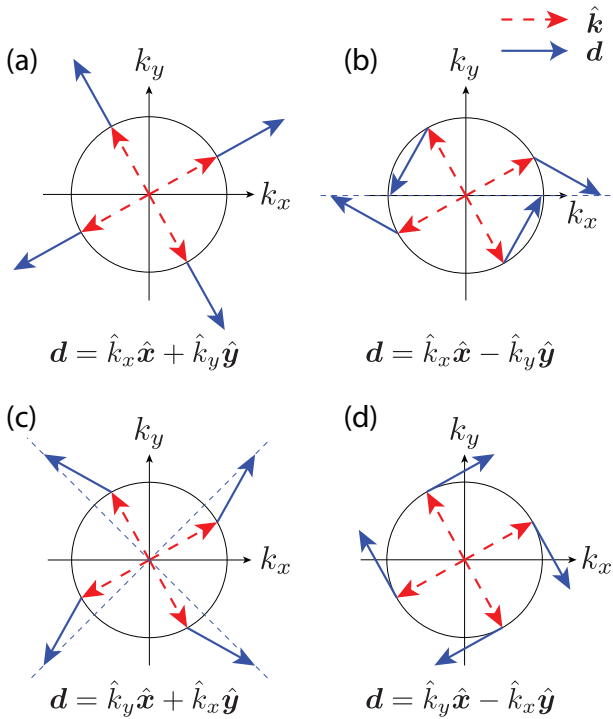}
	\caption{Illustration of the $d$-vectors for the helical states. In terms of $H_{c2,ab}(\phi)$, helical states shown in (a) and (d) are isotropic, while those in (b) and (c)  exhibit four-fold in-plane anisotropies due to the Pauli paramagnetic effect and strong spin-orbit coupling.}
	\label{fig:d-vectors}
\end{figure}

The possible spin-triplet $p$-wave states for Sr$_2$RuO$_4$ are limited by the tetragonal crystal structure with two-dimensional square lattice point group symmetry $C_{4v}$ to the six degenerate states with the $d$-vectors $\hk_x\hx\pm \hk_y\hy$, $\hk_y\hx\pm\hk_x\hy$ and $(\hk_x\pm i\hk_y)\hz$~\cite{Sigrist:1991kj, Rice:1999fn}. The two chiral states $\vd=(\hk_x\pm i\hk_y)\hz$ with $\vd\parallel \hc$ are believed to be stabilized near ${\bm H}=0$~\cite{Mackenzie:2003zz, Maeno:2012ew}, while  with $H\sim H_{c2,ab}$, only the four helical states with $d$-vectors lying in the basal plane could be consistent with the in-plane $H_{c2,ab}$ measurements~\cite{Kittaka:2009hs} by including the effects of Pauli limiting~\cite{Choi:2010ea}. Contrary to the the assumption of very weak spin-orbit coupling, allowing the $d$-vector to rotate to a direction perpendicular to ${\bm H}$, that was argued to explain the Knight shift observations for both ${\bm H}||\hc$ and ${\bm H}\perp\hc$, sufficiently strong spin-orbit coupling  should be assumed to allow for Zeeman energy splitting in spin-triplet pairing states\cite{Rozbicki}. In this case, the degeneracy in the four helical states is lifted~\cite{Sigrist:2005gr}, since each state responds differently to ${\bm H}$, as illustrated in figure~\ref{fig:d-vectors}, two of them manifesting themselves by showing  intrinsic four-fold in-plane anisotropies of $H_{c2,ab}(\phi)$ --- a novel scenario other than earlier postulations of a multi-component order parameter~\cite{Agterberg:2001hh} or the possible misalignment of two domains in the sample~\cite{Kittaka:2009hs}. In this paper, we will calculate the full angular and $T$ dependencies of $H_{c2}(\theta,\phi,T)$ for the four helical states to try to set further restrictions on the possible pairing symmetries in Sr$_2$RuO$_4$.

\section{Model}
\begin{figure}[htbp]
	\centering
	\includegraphics[width=0.45\textwidth]{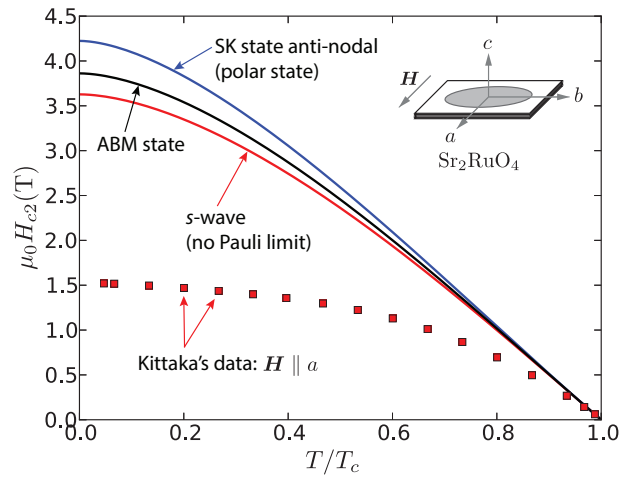}
	\caption{Fits of the chiral ABM,  SK and $s$-wave (without Pauli limiting) states  to the in-plane $H_{c2,a}(T)$ measurements of Sr$_2$RuO$_4$~\cite{Kittaka:2009hs}. The in-plane $H_{c2,ab}(T)$ is strongly suppressed at low temperatures from than predicted from the orbital pairbreaking in these states. Note that in the anti-nodal direction, $H_{c2,ab}(T)$ of the chiral SK state has a first-order transition to that of the nonchiral polar state $\vect{d}=\Delta_0k_x\hz$~\cite{Zhang:2014ul}.}
	\label{fig:noPauli-fit}
\end{figure}

The Fermi surface of Sr$_2$RuO$_4$ consists of three sheets: a quasi-two-dimensional $\gamma$ band, and a pair of quasi-one-dimensional ($\alpha$, $\beta$) bands~\cite{Bergemann:2000ch}. Although still under debate~\cite{Raghu2010aa,Annett:2002ej,Zhitomirsky:2001gb}, the cylindrical $\gamma$ band is widely considered to be the primary source of $p$-wave pairing~\cite{Maeno:2012ew, Huo:2013it}. The small $c$-axis dispersion in this nearly cylindrical $\gamma$ Fermi surface can be incorporated by treating it as an elongated uniaxial ellipsoid, characterized by the effective mass anisotropy of the quasi-particles $m_c\gg m_a=m_b=m_{ab}$. The primary pair-breaking effects established in superconductivity fall into two categories: 1.\ the orbital effect arising from the competition between the coherence of two quasi-particles in a Cooper pair and their individual orbital motions in a magnetic field, i.e., the Landau levels governed by the effective vector potential $\vect{A}$~\cite{Helfand:1966ka}; 2.\ the paramagnetic effect due to the Zeeman energy gained from the interactions between their spins and the field~\cite{Clogston:1962cm}. Highly anisotropic Zeeman interactions are expected in the layered compound Sr$_2$RuO$_4$, described here by an effective diagonal $g$-tensor $g=\mathrm{diag}(g_a,g_b,g_c)$ with $g_{c}\ne g_{a}=g_{b}=g_{ab}$ as in $-\mu_B\vect{S}\cdot g\cdot\overline{m}_{1/2}\cdot\vect{B}$, where $\mu_B$ is the Bohr magneton, $\vect{S}$ is the electron spin, $\overline{m}_{1/2}=\mathrm{diag}(\overline{m}_a^{1/2}, \overline{m}_b^{1/2}, \overline{m}_c^{1/2})$ is the diagonal tensor of the square roots of the relative effective masses $\overline{m}_{\mu}=m_{\mu}/m$ $(\mu=a,b,c)$ with geometric mean effective mass $m=(m_am_bm_c)^{1/3}$, and the magnetic induction $\vect{B}=\mu_0\vect{H}+\vect{M}=\nabla\times\vect{A}$, where  $\vect{M}$ is the magnetization proportional to $\vect{H}$ for the non-ferromagnetic superconductor Sr$_2$RuO$_4$~\cite{Luke:1998bo}. If the $d$-vector is along the $c$ axis, neither the chiral Anderson-Brinkman-Morel (ABM) state $\Delta_0(\hk_x+i\hk_y)\hz$~\cite{Anderson:1961ev, Anderson:1973ee} nor the Scharnberg-Klemm (SK) state $[\Delta_{0+}(\hk_x+\hk_y)+\Delta_{0-}(\hk_x-i\hk_y)]\hz$~\cite{Klemm:1980eh,Zhang:2014ul}, nor generalizations of them obtained by setting $\hk_x\rightarrow\sin(k_xa)$ and $\hk_y\rightarrow\sin(k_ya)$ \cite{Firmo}, could fit the in-plane $H_{c2,ab}$ measurements; for comparison, even the conventional $s$-wave state without Pauli limiting has $H_{c2}(T)$ well above the experimental data of Kittaka {\it et al.} for ${\bm H}||\ha$ (figure \ref{fig:noPauli-fit}). Instead, the helical states have a chance for  Pauli limiting to play the crucial role in suppressing $H_{c2,ab}(T)$, as long as ${\bm H}$ cannot cause the $d$-vectors to rotate.

Hence we model Sr$_2$RuO$_4$ as a clean homogeneous weak-coupling type-II superconductor. Since close to $H_{c2}$,  $\Delta({\bm R}) = \sum_{n=0}^{\infty}a_n\ket{n({\bm R})}$ for the vortex lattice in the mixed state, is constructed from the harmonic oscillator states $\ket{n}$, and  is vanishingly small, the Gor'kov equations for $p$-wave superconductors with a single ellipsoidal Fermi surface can be linearized and transformed to yield~\cite{Klemm:1980eh, Choi:1993cm,Lorscher:2013fx}
\begin{eqnarray}
	\Delta({\bm R}) &=&
	2\pi T N(0)V_0 \sum_{\omega_n} \int \frac{{\rm d}\Omega_{\vk'}}{4\pi} \int_0^{\infty}{\rm d} \xi \exp\bigl[-2|\omega_n|\xi\bigr] \nonumber \\
	&&\times \exp\Bigl[-i\,\mathrm{sgn}\omega_n\xi \vect{v}_F(\hvk')\cdot\Bigl(\alpha\nabla_{\bm R}/i+2e\vect{A}\Bigr)\Bigr] \Delta({\bm R}) \nonumber \\
	&&\times \Bigl(\bigl|\vect{d}(\hvk')\bigr|^2 + [\cos(\alpha g_{\mathrm{eff}}\mu_BB\xi) - 1]\bigl|d_z(\hvk')\bigr|^2\Bigr),
	\label{eq:gap}
\end{eqnarray}
where $N(0)$ is the density of states per spin at the Fermi level, $V_0$ is the pairing amplitude, $\omega_n$ are the fermion Matsubara frequencies, $\vect{v}_F=\vk_F/m$ is the effective Fermi velocity, $\alpha = (\overline{m}_{ab}\sin^2\theta + \overline{m}_c\cos^2\theta)^{1/2}$ characterizes the geometric anisotropy of the Fermi surface, $g_{\mathrm{eff}}=\bigl[g_{c}^2\cos^2\theta'+g_{ab}^2\sin^2\theta'\bigr]^{1/2}$ is the effective $g$-factor experienced by the spins with $\cos\theta'=\sqrt{\overline{m}_c}\cos\theta/\alpha$, $e$ is the electronic charge and the convention $\hbar = c = k_B = 1$ is adopted. We note that the Klemm-Clem (KC) transformations have been performed so that the $\hz'$ direction in (\ref{eq:gap}) is always along ${\bm B}'$~\cite{KlemmClem,Klemm:2011uva,Lorscher:2013fx}.

All of the helical states in figure~\ref{fig:d-vectors} are degenerate in terms of the KC transformed $\bigl|\vect{d}(\hvk')\bigr|^2=(\hk_x'\cos\theta'+\hk_z'\sin\theta')^2+\hk_y'^2$.  However, the $z$-components $\bigl|d_z(\hvk')\bigr|^2=|\hB'\cdot\vect{d}(\hvk')|^2$, which contribute to the Zeeman energy, are distinct for each of the four helical states.
For the helical state $\vect{d}=\hk_x\hx-\hk_y\hy$ in figure~\ref{fig:d-vectors}(b), the KC transformed
\begin{eqnarray}
\bigl|d_z(\hvk')\bigr|^2&=&\bigl(g_{ab}/g_{\mathrm{eff}}\bigr)^2\sin^2\theta'
\bigl[(\hk_x'\cos\theta'+\hk_z'\sin\theta')\cos2\phi\nonumber\\
& &-\hk_y'\sin2\phi\bigr]^2, \label{dz} \end{eqnarray}
is anisotropic in the basal plane\cite{KlemmClem,Klemm:2011uva},
where  $\phi'=\phi$ for $m_a=m_b$, and for consistency we set $\hk_y'\rightarrow\hk_x'$ and $\hk_x'\rightarrow-\hk_y'$.   $\bigl|d_z(\hvk')\bigr|^2$ in state (c) is obtained from that of helical state (b) in (\ref{dz}) by letting $\phi\to\phi-\pi/4$, while  $\bigl|d_z(\hvk')\bigr|^2$ for the helical states (a) and (d) are  respectively obtained by setting $\phi=0$ and $\phi=\pi/4$ in (\ref{dz}). These latter two helical states are therefore isotropic in the basal plane. Accordingly, the helical state (b) with $\vect{d}=\hk_x\hx-\hk_y\hy$ can be used to present the formulation.

We introduce the dimensionless quantities $t=T/T_c(0)$, $b_{c2}=B_{c2}/B_0$ and for the $g$-tensor (via its elements) $\bar{g}=g/g_0$, where $T_c(0)=(2e^C\omega_{\mathrm{c}}/\pi)e^{-1/N(0)V_0}$ is the superconducting transition temperature in zero field, $C=0.577$ is the Euler constant, $\omega_{\mathrm{c}}$ is the energy cutoff from the BCS theory, $B_0=[2\pi T_c(0)]^2/2ev_F^2$ and $g_0=2\pi T_c(0)/\mu_BB_0$. Equation~(\ref{eq:gap}) can be expanded as~\cite{Klemm:1980eh}
\begin{eqnarray}
\!\!\!\!	\bigl[-\ln t + \alpha_n^{(p)} + \alpha_n^{(a)}\bigr]a_n + \beta_{n-2}^{(+)}a_{n-2} + \beta_{n}^{(-)}a_{n+2}& =& 0.\>\>
	\label{eq:recursive}
\end{eqnarray}
The upper critical field $b_{c2}$ is embedded in the coefficients
\begin{eqnarray}
	 \alpha_n^{(p)} &=& \int_0^{\infty}{\rm d}\psi\frac{\sinh^2\psi}{\cosh^3\psi}\int_0^{\infty}{\rm d}\rho\frac{t\sqrt{2/\alpha b_{c2}}}{\sinh\bigl(t\sqrt{2/\alpha b_{c2}}\rho\cosh\psi\bigr)} \nonumber \\
	&&\times \Bigl[e^{-\frac12\rho^2}L_n^{(0)}(\rho^2)F^{(p)}-\sin^2\theta'\Bigr], \\
	 \alpha_n^{(a)}& =& \int_0^{\infty}{\rm d}\psi\frac{1}{2\cosh^3\psi}\int_0^{\infty}{\rm d}\rho\frac{t\sqrt{2/\alpha b_{c2}}}{\sinh\bigl(t\sqrt{2/\alpha b_{c2}}\rho\cosh\psi\bigr)} \nonumber \\
	&&\times \Bigl[e^{-\frac12\rho^2}L_n^{(0)}(\rho^2)F^{(a)}-(1+\cos^2\theta')\Bigr], \\
	\beta_n^{(\pm)} &=& \int_0^{\infty}{\rm d}\psi\frac{1}{4\cosh^3\psi}\int_0^{\infty}{\rm d}\rho\frac{t\sqrt{2/\alpha b_{c2}}}{\sinh\bigl(t\sqrt{2/\alpha b_{c2}}\rho\cosh\psi\bigr)} \nonumber \\
	&&\times e^{-\frac12\rho^2}\frac{-\rho^2}{\sqrt{(n+2)(n+1)}}L_{n}^{(2)}(\rho^2) \nonumber \\
	&&\times \Bigl[-\sin^2\theta'+G(\cos^2\theta'\cos^2 2\phi-\sin^2 2\phi)\nonumber\\
 & &\>\>\pm iG\cos\theta'\sin4\phi\Bigr],
\end{eqnarray}
where the $L_{n}^{(m)}$ are the associated Laguerre polynomials, and
\begin{eqnarray}
	F^{(p)}&=& \sin^2\theta'+G\sin^2\theta'\cos^2 2\phi, \\
	F^{(a)}&=& 1+\cos^2\theta'+G(\cos^2\theta'\cos^{2}2\phi+\sin^{2}2\phi)
\end{eqnarray}
with $G = \bigl[\cos\bigl(\bar{g}_{\mathrm{eff}}\sqrt{\alpha b_{c2}/2}\bigr)-1\bigr]\bigl(\bar{g}_{ab}/\bar{g}_{\mathrm{eff}}\bigr)^2\sin^2\theta'$. The solution to (\ref{eq:recursive}) constitutes the determinant of the (infinite order) tridiagonal matrix constructed from the coefficients of $a_n$ ($n=0,2,\dots$), which can be solved numerically for arbitrary $t$. To calculate $B_{c2}=\mu_0H_{c2}$ for non-magnetic superconductors, usually the first $3$ or $4$ orders produce sufficiently accurate results to show all of the essential features.

\section{Results}

\begin{figure}[htbp]
	\centering
	\includegraphics[width=0.45\textwidth]{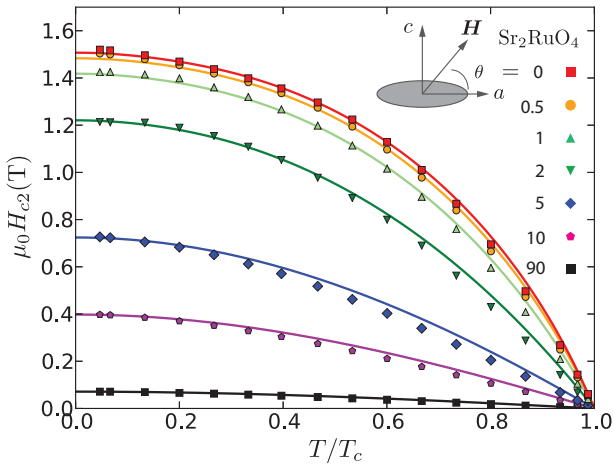}
	\caption{Fits to the angular dependent $H_{c2,a}(\theta,T)$ measurements of Sr$_2$RuO$_4$ sample~\cite{Kittaka:2009hs} using  helical state (b) with $\vect{d}=\hk_x\hx-\hk_y\hy$ with $\bar{g}_{ab}=1.9$, $\bar{g}_{c}=0.2$.}
	\label{fig:Pauli-fit}
\end{figure}

Figure~\ref{fig:Pauli-fit} shows our fits to the angular dependent $H_{c2,a}(\theta,T)$ measurements of Kittaka {\it et al.} on a sample of Sr$_2$RuO$_4$ ($T_c(0)=1.503$ K)~\cite{Kittaka:2009hs} using  helical state (b) with $\vect{d}=\hk_x\hx-\hk_y\hy$. The appropriateness of an elongated uniaxial ellipsoidal Fermi surface for the $\gamma$ band is verified by the huge effective mass anisotropy $m_c/m_{ab}=1067$ estimated from the slopes of $H_{c2,a}(T)$ at $T_c(0)$ in the $[100]$ and $[001]$ crystal directions where Pauli limiting effects are neglibible. Down to low $t$, a suitable choice of the effective $g$-factor will further suppress the $H_{c2}$ curves, especially for those with $\theta<5^\circ$ (c.f.\ figure~\ref{fig:noPauli-fit}). Although the $H_{c2,a}(T,\theta>5^\circ)$ data appear to follow the anisotropic effective mass model~\cite{Kittaka:2009hs,Zhang:2014ul,Lorscher:2013fx}, one should nevertheless take into consideration the intrinsic anisotropy of $H_{c2}(\theta)$ raising from the point nodal structures of the helical states ($[H_{c2,ab}/H_{c2,c}]_{T\to T_c(0)}=\sqrt{2}$ for an isotropic Fermi surface)~\cite{Zhang:2014ul}. For an overall best fit, the effective $g$-tensor was evaluated to have the diagonal elements $\bar{g}_{c}=0.2$ and $\bar{g}_{ab}=1.9$. Obviously, the small-valued $\bar{g}_{c}$ doesn't contribute to $H_{c2,c}$ since $\vect{d}\perp \hc$ for the helical states, but it plays a role in determining $H_{c2}(\theta)$ for $(0^\circ<\theta<90^\circ)$.

\begin{figure}[htbp]
	\centering
	\includegraphics[width=0.45\textwidth]{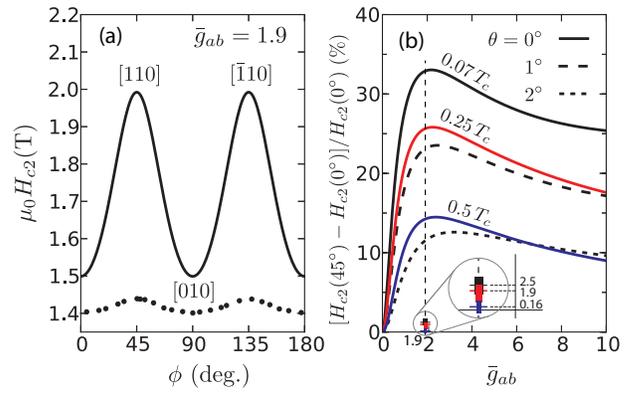}
	\caption{In-plane $H_{c2,ab}(\phi)$ anisotropy of  helical state (b) with $\vect{d}=k_x\hx-k_y\hy$. (a) $H_{c2,ab}(\phi)$ with an effective $g$-factor $\bar{g}_{ab}=1.9$ at $0.13$~K . The amplitude of the predicted anisotropy (solid) is an order of magnitude larger than that (dotted curve) observed in Sr$_2$RuO$_4$ by Mao {\it et al.}\cite{Mao:2000gn}. (b) Effects of  $g_{\rm eff}$ to the relative magnitudes of the in-plane anisotropy at various temperatures and field misalignments. Anisotropies comparable to the experiments only occur with small $\overline{g}_{ab}$ values.  The symbols at the bottom represent the data of Kittaka {\it et al.}\cite{Kittaka:2009hs}.}
	\label{fig:anisotropy}
\end{figure}

We remark that all the helical states listed in figure~\ref{fig:d-vectors} could  equally well fit the data shown in figure 3, as the  differences in their $H_{c2}$ values only appear in their in-plane ($\phi$) anisotropies. As seen from (\ref{eq:gap}), in the absence of Pauli limiting, $H_{c2,ab}(\phi)$ for the helical states are isotropic in the basal plane. However, with the fitting parameter $\bar{g}_{ab}=1.9$, the $H_{c2,ab}(\phi)$ at $0.13$~K for the helical states (b) and (c) in figure~\ref{fig:d-vectors} exhibit four-fold in-plane azimuthal anisotropies with a relative amplitude as large as $30\%$ (figure~\ref{fig:anisotropy}(a)) and a phase shift of $\pi/4$ between them, while those for states (a) and (d) remain isotropic in the $ab$ plane. The observed in-plane anisotropy of $H_{c2,ab}(\phi)$ is at most $3\%$ and  disappears either above $0.8$ K or with a field misalignment of less than $1^\circ$~\cite{Kittaka:2009hs,Mao:2000gn}. The calculated anisotropy for helical state (b) with $\vect{d}=\hk_x\hx-\hk_y\hy$ state persists for $T > T_c/2$ and for field misalignments greater than $2^\circ$ (figure~\ref{fig:anisotropy}(b)).  Thus, this parallel-spin $p$-wave state can explain the strong Pauli limiting for ${\bm B}\perp\hc$, but the details are not in precise agreement with the experimental observations~\cite{Kittaka:2009hs,Mao:2000gn}.

\section{Discussion}

A multi-component order parameter proposed to  interpret the in-plane $H_{c2,ab}(\phi)$ anisotropy in reference~\cite{Mao:2000gn} turns out to have a similar problem of a large magnitude of  the in-plane anisotropy~\cite{Agterberg:2001hh}. There could also be two slightly misaligned crystals in the same sample~\cite{Kittaka:2009hs}, and the smaller region of the hysteretic magnetization data below 0.8 K in the more recent data of Yonezawa {\it et al.} than in the older Mao {\it et al.} and Deguchi {\it et al.} data are consistent with this scenario\cite{Deguchi:2002bv,Kittaka:2009hs,Yonezawa:2013je,Mao:2000gn}.  Others think that this first-order transition below 0.8 K is more intrinsically due to a Fulde-Farrell-Larkin-Ovchinnikov state, entered below 0.55$T_c$ (close to $0.8$~K in Sr$_2$RuO$_4$)~\cite{Matsuda:2007aa}. Based on the present calculations, if the Pauli pair-breaking effect is demanded as the source for the suppression on $H_{c2,ab}$,  helical state (b) with $\vect{d}=\hk_x\hx-\hk_y\hy$ has the same four-fold anisotropy with the same phase as in the experiments. Helical state (c) with $\vect{d}=\hk_y\hx+\hk_x\hy$ has the four-fold anisotropy differing in phase by $\pi/4$.  However, both of these  azimuthal anisotropies  are much stronger than that observed in experiment.  However, the other helical (a) and (d) $p$-wave states with $\vect{d}=\hk_x\hx+\hk_y\hy$ and $\hk_y\hx-\hk_x\hy$  are predicted to have no azimuthal anisotropies at all. Including $ab$-planar anisotropy on the $\gamma$ Fermi surface could lead to a small azimuthal anisotropy of $H_{c2}(90^{\circ},\phi,T)$, but normally Fermi surface anisotropy is largest near to $T_c$.  Thus, a single purported triplet-spin order parameter for  Sr$_2$RuO$_4$ is still elusive.  We note, however, that there are many examples in which the Knight shift observations have been misleading and/or are also in apparent conflict with the upper critical field results~\cite{HallKlemm,Noer:1964aa}, strongly suggesting that a new theory of the Knight shift might lead to a possible resolution of the symmetry of the order parameter in Sr$_2$RuO$_4$\cite{Mazin,HallKlemm}.

In summary, we studied the four helical $p$-wave states potentially realized in Sr$_2$RuO$_4$ at $H_{c2}$ by fitting the angular dependent $H_{c2,a}(\theta,\phi,T)$ measurements, taking the Pauli paramagnetic effects into account by imposing strong spin-orbit coupling effects as the origin of the $H_{c2,ab}$ suppression. In the ranges of the fitting parameters, one of the four helical states was predicted to have in-plane $H_{c2}(90^{\circ},\phi,T)$ four-fold azimuthal anisotropy with the same phase as observed, but both that azimuthal anisotropy and that from the (c) helical  state with the anisotropy shifted by $\pi/4$ in phase, had amplitudes that were predicted to be   much stronger than that observed in Sr$_2$RuO$_4$. The $H_{c2}(90^{\circ},\phi,T)$ behaviors of the two other helical $p$-wave states were predicted  to be completely independent of $\phi$, as long as in-plane Fermi surface anisotropy could be safely ignored. Other attempts to fit an order parameter such as $\Delta_0[\sin(k_xa)+i\sin(k_ya)]$ with the low-$T$ specific heat $C_V\sim T^2$ dependence failed to confront the very strong Pauli limiting of $H_{c2}(90^{\circ},\phi,T)$\cite{Firmo}.   Thus, the thermodynamic zero-field specific heat measurements appear to be in direct conflict with the field-dependent thermodynamic specific heat and magnetization measurements of the upper critical field\cite{Deguchi:2002bv,Kittaka:2009hs,Mao:2000gn,Yonezawa:2013je}. Further calculations to try to fit the excellent  scanning tunneling microscopy  results of Suderow {\it et al.} with a $p$-wave order parameter are also needed\cite{Suderow}.  A point node in a helical $p$-wave order parameter might smear the sharp density of states walls they observed, but an accurate calculation is needed to quantify this possible disagreement.

\section*{Acknowledgments}
We thank S. Kittaka for supplying us with his published $H_{c2}(\theta,\phi,T)$ data.
This work is partially supported by the National Natural Science Foundation of China (Grant No.~11274039) and the Specialized Research Fund for the Doctoral Program of Higher Education (No.~20100006110021). QG acknowledges helpful discussions with Prof.~\mbox{John Chalker} and is grateful for the support from the China Scholarship Council and the hospitality of the Rudolf Peierls Centre for Theoretical Physics, University of Oxford.

\section*{References}


\end{document}